\documentclass[onecolumn,amsmath,amssymb]{qm2008_abs}
\usepackage{graphicx}
\usepackage{dcolumn}
\usepackage{bm}
\begin{document}

\title{{\Large  Beam-Energy and System-Size Dependence of Dynamical Net Charge Fluctuations }}

\bigskip
\bigskip
\author{\large Monika Sharma for the STAR Collaboration}
\email{monika@rcf.rhic.bnl.gov} \affiliation{Department of Physics
and Astronomy, Wayne State University, Detroit, Michigan, USA}
\bigskip
\bigskip

\begin{abstract}
\leftskip1.0cm \rightskip1.0cm We present measurements of net
charge  fluctuations in $Au + Au$ collisions at $\sqrt{s_{NN}} = $
19.6, 62.4, 130, and 200 GeV, $Cu + Cu$ collisions at
$\sqrt{s_{NN}} = $ 62.4, 200 GeV, and $p + p$ collisions at
$\sqrt{s} = $ 200 GeV using the net charge dynamical
fluctuations measure  $\nu_{+-,dyn}$. The dynamical fluctuations
are non-zero at all energies and exhibit a rather modest
dependence on beam energy. We find that at a given energy and collision 
system, net charge dynamical fluctuations violate $1/N_{ch}$ scaling, 
but display approximate $1/N_{part}$ scaling. We observe strong dependence 
of dynamical fluctuations on the azimuthal angular range and pseudorapidity
widths.
\end{abstract}
\maketitle
\section{Introduction}
Anomalous transverse momentum and net charge event-by-event
fluctuations have been proposed as indicators of the formation of
a quark gluon plasma (QGP) in the midst of high-energy heavy ion
collisions. Entropy conserving hadronization of a plasma of quarks
and gluons could produce a final state characterized by a dramatic
reduction of the net charge fluctuations relative to that of a
hadron gas \cite{JeonKoch00}. Measurements of net charge
fluctuations were reported by both PHENIX \cite{AdcoxPRC89} and
STAR \cite{Adams03c} collaborations on the basis of $Au + Au$ data
acquired during the first RHIC run at $\sqrt{s_{NN}} =$ 130~GeV.
In this work, we report a systematic study of the net charge
fluctuations as a function of the system size and their energy
using $\nu_{+-{\rm,dyn}}$ fluctuation measure, defined as follows:
\begin{equation}
\nu _{ +  - ,dyn}  = \frac{{\left\langle {N_ +  (N_ +   - 1)} \right\rangle }}{{\left\langle {N_ +  } \right\rangle ^2 }} + \\
\frac{{\left\langle {N_ -  (N_{-}   - 1)} \right\rangle
}}{{\left\langle {N_ -  } \right\rangle ^2 }} -
2\frac{{\left\langle {N_ -  N_ +  } \right\rangle }}{{\left\langle
{N_ -  } \right\rangle \left\langle {N_ +  } \right\rangle }}
\end{equation}
where $N_{\pm}$ are the number of positively and negatively
charged particles in the acceptance of interest. We present
measurements of dynamical net charge fluctuations in $Au + Au$
collisions at $\sqrt{s_{NN}} = $ 19.6, 62.4, 130, and 200 GeV, $Cu
+ Cu$ collisions at $\sqrt{s_{NN}} = $ 62.4, 200 GeV and in $p + p$
collisions at $\sqrt{s_{NN}} = $ 200 GeV. We study the beam energy,
system size and collision centrality dependencies quantitatively
in order to identify possible signature of the formation of a QGP.
\section{Experimental Method}
The data used in this analysis were measured using the Solenoidal
Tracker at RHIC (STAR) detector during the 2001, 2002, 2004 and
2005 data RHIC runs at Brookhaven National Laboratory. The $Au +
Au$ collisions at $\sqrt{s_{NN}} =$ 62.4, 130, and 200 GeV data
and $Cu + Cu$ collisions at $\sqrt{s_{NN}} =$ 62.4 and 200 GeV
were acquired with minimum bias triggers. For 19.6 GeV data, a
combination of minimum bias and central triggers was used.
Technical descriptions of the STAR detector and its components are
published in technical reports \cite{Star03}.

This analysis used tracks from the TPC with transverse momentum in
the range $0.2  < p_T < 5.0$ GeV/c with pseudorapidity $|\eta| <
0.5$. The centrality bins were calculated as a fraction of this
multiplicity distribution starting at the highest multiplicities.
The ranges used were 0-5\% (most central collisions), 5-10\%,
10-20\%, 20-30\%, 30-40\%, 40-50\%, 50-60\%, 60-70\%, and 70-80\%
(most peripheral) for $Au + Au$ collisions. Similarly, collision
centrality slices used in $Cu + Cu$ collisions are 0-10\% (most
central), 10-20\%, 20-30\%, 30-40\%, 40-50\% and 50-60\% (most
peripheral). Each centrality bin is associated with an average
number of participating nucleons, $N_{part}$, using Glauber Monte
Carlo calculation \cite{StarGlauber03}.
\section{Net Charge Fluctuation Results}
\begin{figure}[!htb]
\centering
\resizebox{7.5cm}{5.5cm}{\includegraphics{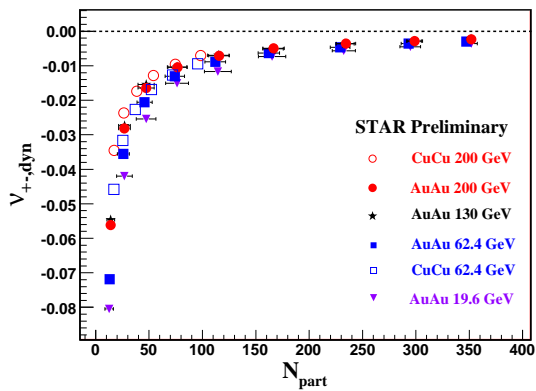}}
\caption[]{Net charge dynamical fluctuations, $\nu_{+-{\rm,dyn}}$,
of particles produced within pseudorapidity $|\eta|<0.5$, as
function of the number of participating nucleons.} \label{fig1}
\end{figure}

We present, in Fig. \ref{fig1}, measurements of the dynamical net
charge fluctuations, $\nu_{+-{\rm,dyn}}$, as a function of
collision centrality in $Au + Au $ collisions at $\sqrt{s_{NN}} =$
19.6, 62.4, 130, and 200 GeV, $Cu + Cu$ collisions at
$\sqrt{s_{NN}} =$ 62.4 and 200 GeV.
The dynamical net charge fluctuations, in general, exhibit a
monotonic dependence on the number of participating nucleons. At a
given number of participants the measured fluctuations also
exhibit a modest dependence on beam energy, with
$\nu_{+-{\rm,dyn}}$ magnitude being the largest in $Au + Au$
collisions at $\sqrt{s_{NN}} = $~19.6~GeV. The $\nu_{+-{\rm,dyn}}$
values measured for $p + p$ collisions at $\sqrt{s} = $ 200
GeV amounts to -0.230 $\pm$ 0.019(stat). Dependence on the 
number of participating nucleons can be seen in Fig. \ref{fig2}(b), 
discussed in Section III A.
\subsection{Collision Centrality Dependence}
\begin{figure}[htb]
    \begin{minipage}{0.4\textwidth}
    \centering
    \includegraphics[width=7.5cm]{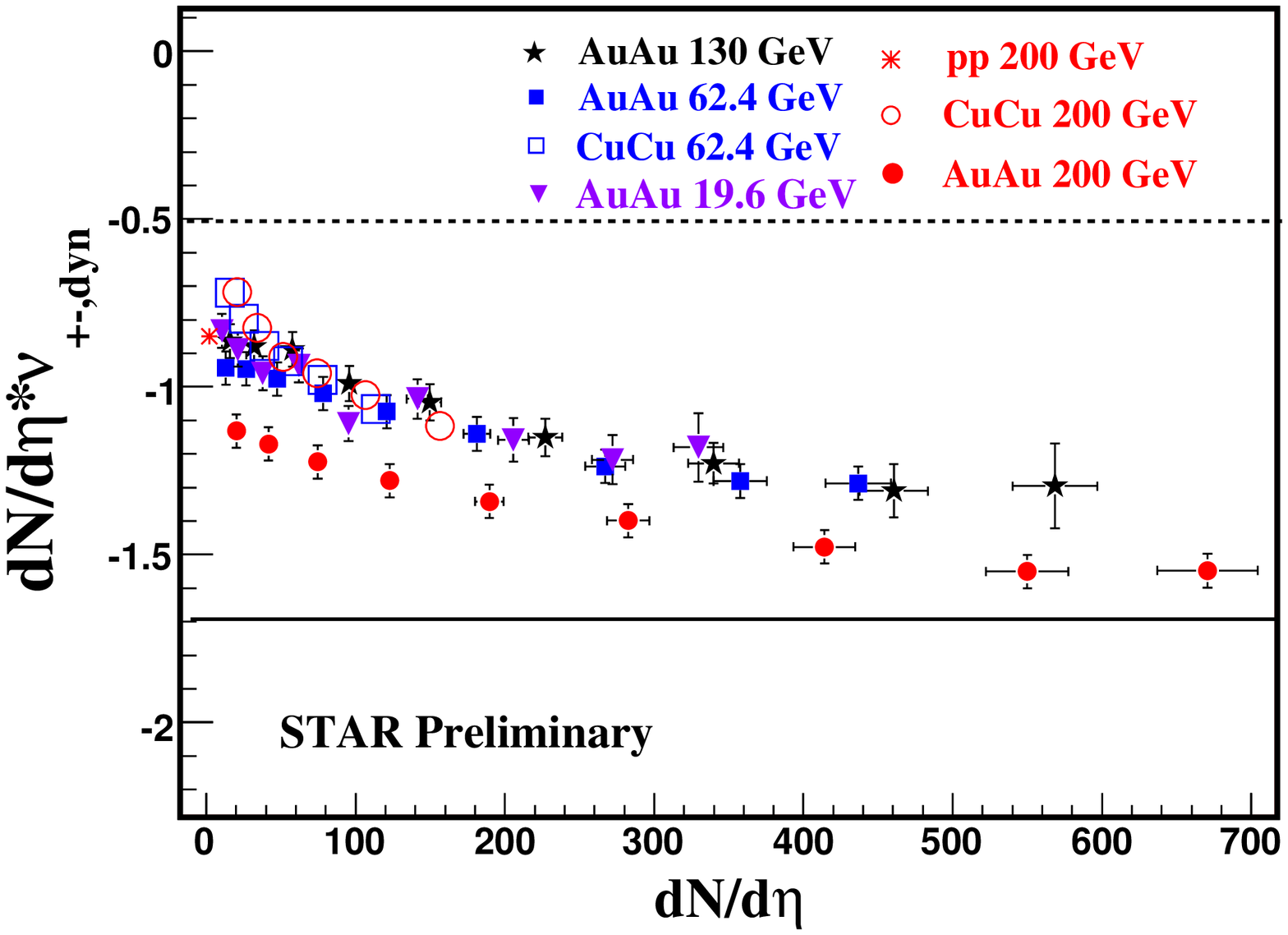}
   \end{minipage}
   \hspace{1.0cm}
   \begin{minipage}{0.4\textwidth}
   \centering
    \includegraphics[width=7.3cm]{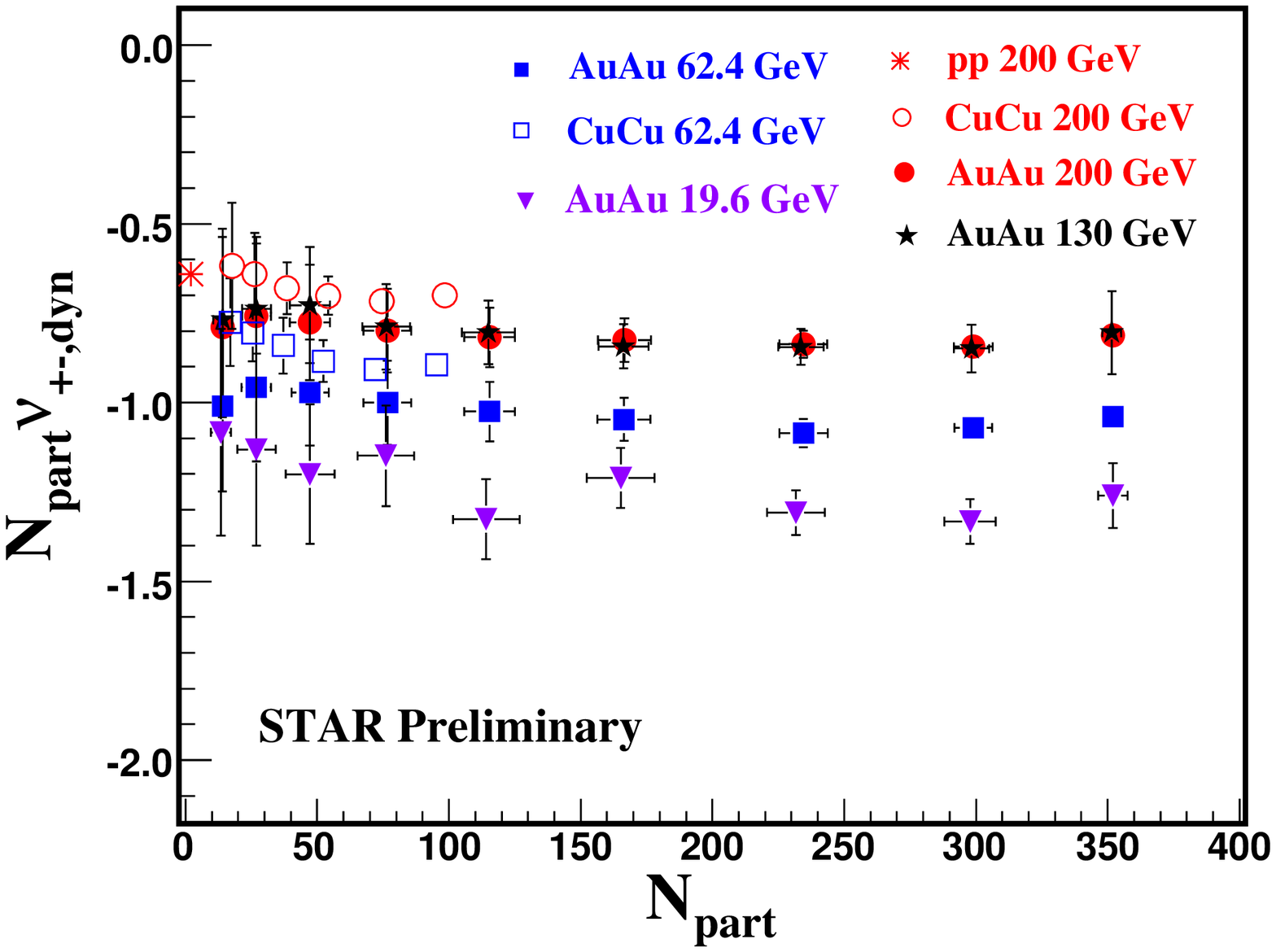}
   \end{minipage}
   \caption[]{Net charge dynamical fluctuations, $\nu_{+-{\rm,dyn}}$,
of particles produced with pseudorapidity $|\eta|<0.5$ scaled by
(a) the multiplicity, $dN_{ch}/d\eta$, (b) the number of
participants. The dashed line corresponds to charge conservation
effect and the solid line to the prediction for a resonance gas.}
   \label{fig2}
\end{figure}

Fig. \ref{fig2}(a) shows the dynamical fluctuations scaled by the
measured particle multiplicity density in pseudorapidity space
($dN_{ch}/d\eta$). Data from $Au + Au$ and $Cu + Cu$ collisions
are shown with solid and open symbols, respectively. Values of
$dN_{ch}/d\eta$ used for the scaling correspond to efficiency
corrected charged particle multiplicities measured by STAR
\cite{StardNdeta} and PHOBOS \cite{Phobos01}. The magnitude of
$\nu_{+-{\rm,dyn}}$ scaled by $dN_{ch}/d\eta$ for $Au + Au$ 200
GeV data is significantly different from the results of other systems and energies. 
 The observed $|\nu_{+-{\rm,dyn}}dN_{ch}/d\eta|$ increases with the  
increase in collision centrality. The
 dashed line in the figure corresponds to charge conservation effect
  and the solid line to the prediction for a
resonance gas. The data indicates that dynamical net
charge fluctuations, scaled by $dN_{ch}/d\eta$ for the most central collisions 
are qualitatively consistent with resonance gas prediction \cite{JeonKoch00}.

Fig. \ref{fig2}(b) presents the dynamical fluctuation scaled by
the number of participants, $N_{part} \nu_{+-{\rm,dyn}}$ as a
function of the  number of participants. 
 The measured data scaled by the number of participants
($N_{part}$) are thus consistent with no or a very weak
collision centrality dependence. However, a definite system size
and energy dependence is observed. Vertical error bars represent statistical uncertainties.
\subsection{Longitudinal and Azimuthal Dependencies of the Dynamical Fluctuations}
\begin{figure}[htb]
    \begin{minipage}{0.35\textwidth}
    \includegraphics[width=7.25cm]{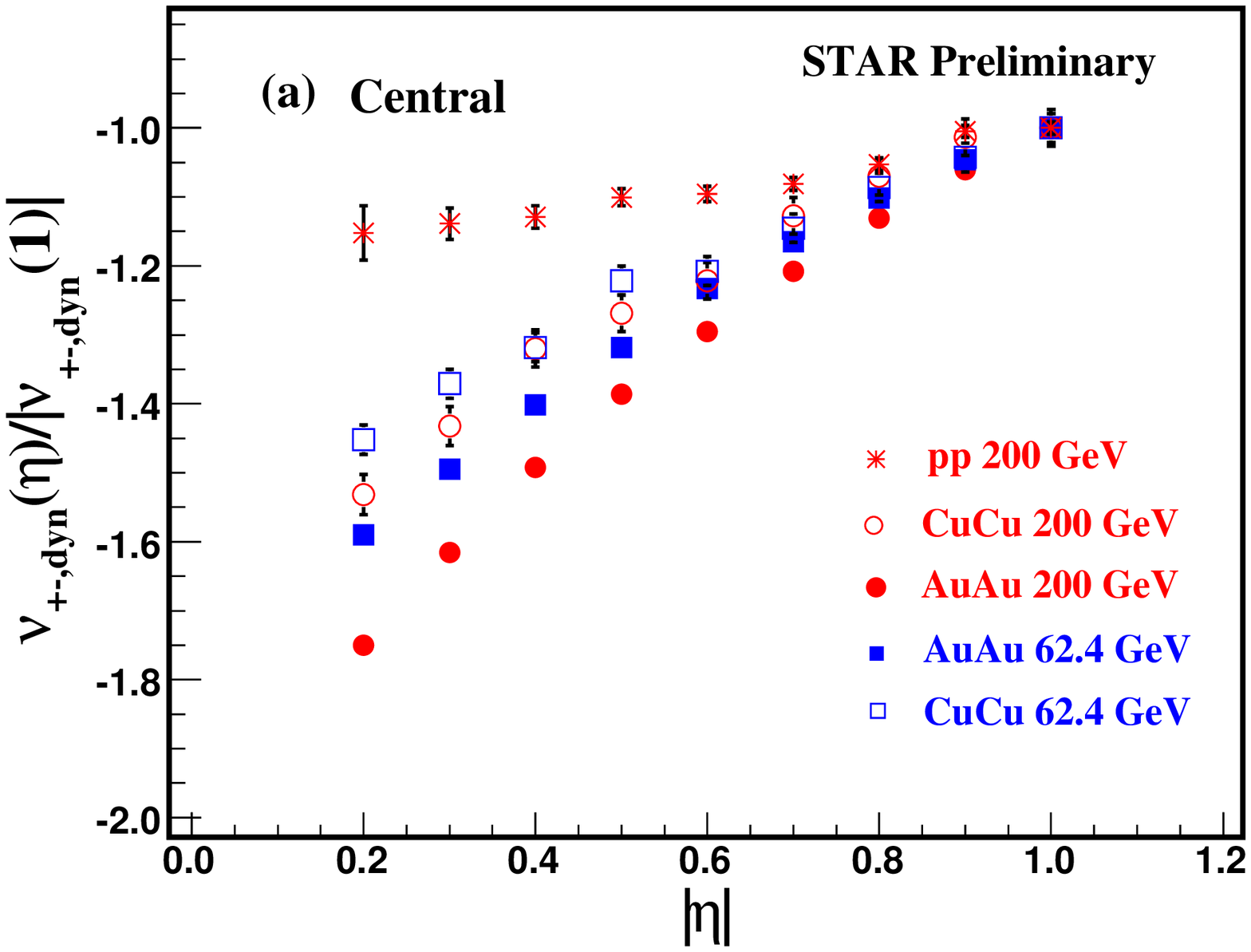}
    \caption []{Dynamical fluctuations $\nu_{+-{\rm,dyn}}$, normalized
to their value for $|\eta|<1$, as function of the integrated
pseudorapidity range. Data for $Au + Au$ collisions at
$\sqrt{s_{NN}}=$ 62.4, 200 GeV (0-5\%) along with data for $Cu +
Cu$ collisions at $\sqrt{s_{NN}}=$ 62.4, 200 GeV (0-10\%), are
compared to inclusive $p + p$ data at $\sqrt{s_{NN}}=$ 200
GeV.\label{fig3}}
   \end{minipage}
   \hspace{2.75cm}
   \begin{minipage}{0.4\textwidth}
   \centering
    \includegraphics[width=7.25cm]{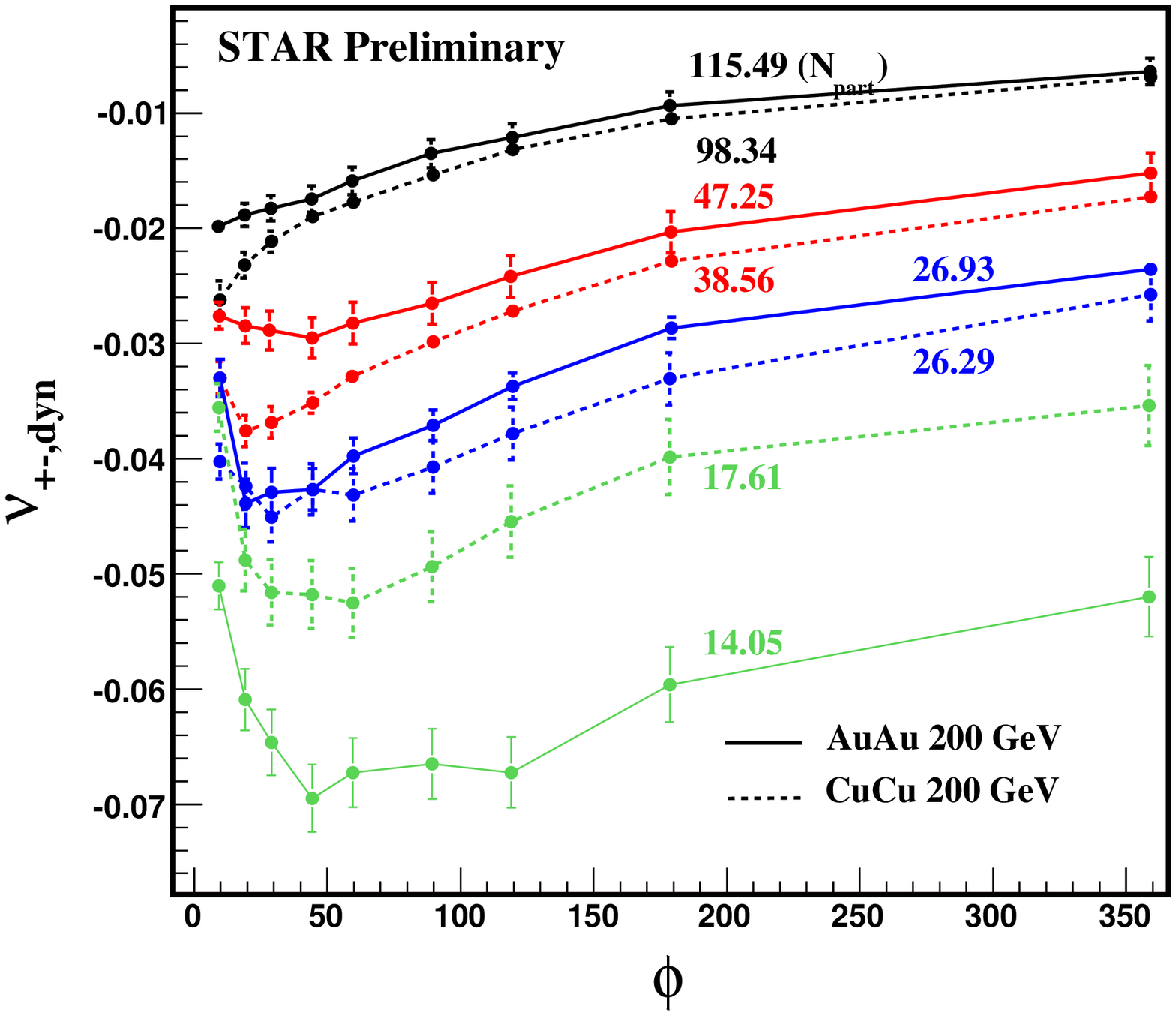}
    \caption{Dynamical fluctuations $\nu_{+-{\rm,dyn}}$, as a function of the
    integrated azimuthal range $\phi$ for similar
 number of participating nucleons for $Au + Au$ 200 GeV and $Cu + Cu$ 200 GeV. \label{fig4}}
   \end{minipage}
\end{figure}

We plot in Fig. \ref{fig3} values of $\nu_{+-{\rm,dyn}}(\eta)$
measured for different ranges of pseudorapidity, $\eta$,
normalized by the magnitude of $\nu_{+-{\rm,dyn}}(\eta)$ for a
pseudorapidity range $|\eta|<1$ ($\nu_{+-{\rm,dyn}}(1)$) to enable
comparison of different centralities, beam energies and system
size. The magnitude of the normalized correlation is maximum for
the smallest pseudorapidity ranges and decreases monotonically to
unity, at all energies and centralities, with increasing
pseudorapidity range. This shows that the collision dynamics  in
$p + p$ collisions, 0-10\% $Cu + Cu$ and 0-5\% $Au + Au$
collisions are significantly different. Indeed, we find the
relative magnitude of the correlations measured for $|\eta|<0.5$
increases by nearly 25\% for $Au + Au$ 200 GeV relative to those
in $p + p$. Note in particular that the slope
($d\nu_{+-{\rm,dyn}}/d\eta$) in $p + p$, $Cu + Cu$ and $Au + Au$
systems depends on the correlation length (in pseudorapidity): the
shorter the correlation, the larger the slope. The observed
distributions then indicate that the correlation length is shorter
for central collisions and for larger systems, in agreement with
the observed reduction of the charge balance function
\cite{StarBalanceFct}.
Fig. \ref{fig4} displays measured values of $\nu_{+-{\rm,dyn}}$ as
a function of azimuthal angle ranges from 10 to 360 degrees for
$Au + Au$ and $Cu +Cu$ collisions at 200 GeV. $Au + Au$ and $Cu +
Cu$ collisions are compared at similar number of participating
nucleons. One observes that the magnitude of $\nu_{+-{\rm,dyn}}$
with respect to azimuthal angle, $\phi$, is similar for similar
number of participating nucleons in both systems with best
agreement for collisions with more than 20 participants.
\section{Summary and Conclusions}
We have presented measurements of dynamical net charge
fluctuations in $Au + Au$ collisions at $\sqrt{s_{NN}} =$ 19.6,
62.4, 130, 200 GeV, $Cu + Cu$ collisions at $\sqrt{s_{NN}} =$
62.4, 200 GeV and $p + p$ collisions at $\sqrt{s_{NN}} =$ 200 GeV,
using the measure $\nu_{+-{\rm,dyn}}$. The fluctuations are non
vanishing at all energies and exhibit a rather modest dependence
on beam energy in the range 19.6 $<$ $\sqrt{s_{NN}}$ $<$ 200 GeV
for $Au + Au$ as well as $Cu + Cu$ collisions. Net charge
dynamical fluctuations violate the trivial $1/N_{ch}$ scaling
expected for nuclear collisions consisting of independent
nucleon-nucleon interactions. Scaled dynamical net charge
fluctuations $|\nu_{+-{\rm,dyn}}dN_{ch}/d\eta|$ grow by up to 40\%
from peripheral to central collisions. This centrality dependence
may arise, in part due to the large radial collective flow
produced in $Au + Au$ collisions. We also studied fluctuations as
a function of azimuthal angle and pseudorapidity and found that
dynamical fluctuations exhibit a strong dependence on rapidity and
azimuthal angular ranges which could be
attributed in part to radial flow effects.\\[0.5cm] 

\end{document}